\documentclass[referee]{aa}
\usepackage{graphics}
\begin{document}

\thesaurus{11.09.3,11.03.3}

\title{EUV and X-ray observations of Abell 2199:
a three-phase intracluster medium with a massive warm component}

\author{Richard \,Lieu\inst{1} \and  Massimiliano \,Bonamente\inst{2}
\and Jonathan P. D.\,Mittaz\inst{3}}

\offprints{R. Lieu}

\institute{Department of Physics, University of Alabama,
Huntsville, AL 35899, U.S.A. 
\and Osservatorio Astrofisico di Catania, Via S. Sofia 78, I-95125 Catania, Italy
\and 
Mullard Space Science Laboratory, UCL,
Holmbury St. Mary, Dorking, Surrey, RH5 6NT, U.K.}

\date{Received \hspace{3cm} / Accepted }
\titlerunning{EUV and X--ray observations of A2199}
\maketitle

\begin{abstract}
Various independent ways of constraining key cosmological parameters yielded
a consensus range of values which indicates that 
at the present epoch the bulk of
the universe's baryons is in the form of a warm
($\sim$ 10$^6$ K) gas - a temperature regime which renders them
difficult to detect.  The discovery of EUV and soft X-ray excess emission
from clusters of galaxies was originally interpreted as the first direct
evidence for the large scale presence of such a warm component.
We present results from an EUVE Deep Survey (DS) observation of the 
rich cluster Abell 2199 in the Lex/B (69 - 190 eV) filter passband.
The soft excess radial trend (SERT), shown by a plot 
against cluster radius $r$ of the percentage
EUV emission $\eta$ observed above
the level expected from
the hot intracluster medium (ICM), behaves as
a simple function of $r$ which decreases monotonically towards
$r = 0$; it smoothly  turns negative
at $r \sim$ 6 arcmin, inwards of this radius 
the EUV is absorbed by cold matter with a
line-of-sight column density of $\geq$ 2.7 $\times$ 10$^{19}$ cm$^{-2}$.  
The centre of absorption is offset from
that of the emission by $\sim$ 1 arcmin, and the area involved is
much larger than that of the cooling flow.  These facts
together provide strong evidence for a centrally 
concentrated but cluster-wide distribution of clumps of cold gas
which co-exist with warm gas of similar spatial properties.
Further, the simultaneous modeling of
EUV and X-ray data requires a warm component even within the region
of absorption.
The phenomenon demonstrates a three phase ICM,
with the warm
phase estimated to be $\sim$ 5-10 times
more massive than the hot.

\end{abstract}
Abell 2199 is a well-studied X-ray bright cluster of galaxies
(e.g. Siddiqui et al. 1998).
In a recent study (Lieu et al. 1999a; LBM99) 
we presented a first 
observation of the cluster with the {\it Extreme Ultraviolet Explorer} 
(EUVE; Bowyer \& Malina 1991);
the cluster showed clear evidence of {\it excess} emission
in the DS Lex/B filter ($\sim$ 65-190 eV passband) above the
thermal contribution from the hot intra-cluster medium (ICM).
The A2199 sky area was re-observed by EUVE for 
$\sim$  57 ksec in February 1999 (see Lieu et al. 1999b; L99). 
The program featured an {\it in situ} background measurement by pointing
at small offset from the cluster, which
asserted the correctness of
the original method of determining (and subtracting) the
background from an asymptotically flat region of
the radial profile.
Complementary data in
the X-ray (0.2 - 2.0 keV) passband, as gathered by
a ROSAT PSPC observation which took place in July 1990, with an exposure
of 8.5 ksec, were 
extracted from the public 
archive~\footnote{See the High Energy Astrophysics Archive
available at http://heasarc.gsfc.nasa.gov/docs/rosat/archive.html. 
Same PSPC observation (RP number 150083) as in L99 and LBM99 was
consulted for the present work.}. 
For correct comparison between the EUV and X-ray emissions, the
Galactic HI column density was measured at
N$_H$ = (8.3 $\pm$ 1.0) $\times$ 10$^{19}$ cm$^{-2}$
by a dedicated observation
at Green Bank (Kaastra et al. 1999), and was found to be spatially smooth.
The EUV and X-ray data were simultaneously modeled with a
thin plasma emission code (Mewe et al. 1985;
Mewe et al. 1986) and appropriate line-of-sight Galactic
absorption (Morrison \& McCammon 1983)
for the above value of N$_H$.  At a given radius the hot ICM
was assumed to be isothermal, with the abundance fixed at
0.3 solar apart from the cooling flow region where the parameter
became part of the data fitting in order to account for any possible
abundance gradient within this region (a different way of handling
the abundance does not sensitively affect the results presented in
this work).

The forementioned model, when applied to the
PSPC spectra at all radii, generally leads to acceptable fits.  
At low energies the EUV measurements gave crucial
new information.  The overall effect is a soft excess as
reported previously.
A plot of the SERT indicates, however,
that the percentage EUV excess at a given radius is
less at the centre.  In fact, the trend takes
the form of a negative central excess (i.e. absorption, see Fig. 1), 
which
steadily rises with radius until the
6 arcmin point, beyond that the fractional excess turns positive and
continues to increase until the limiting radius of EUV
detection ($\sim$ 20 arcmin, Lieu et al. 1999b).  In the
present {\it paper} we interpret the
results physically, and demonstrate that such a trend,
together with the spectral data, provide compelling
reasons for a three phase model of the ICM, with the warm
intermediate phase as origin for the soft excess.

We first address the outer parts of the cluster, where the data
present formidable difficulties to the non-thermal interpretation
of the soft excess which postulates a large population of
intracluster relativistic electrons undergoing inverse-Compton (IC)
interaction with the cosmic microwave background (CMB, see
Ensslin \& Biermann 1999, Sarazin \& Lieu 1998).  
Fig. 2 shows a composite plot of 
the EUV and X-ray data for the 12 - 15 arcmin annulus.  The
prominent EUV excess, unaccompanied by any similar effect in
soft X-rays, implies that the bulk of the relativistic  electrons
have energies below 200 MeV, a cut-off which is most
obviously understood as due to aging (i.e. synchrotron and
inverse-Compton losses): the electrons are at
least 3 $\times$ 10$^9$ years old.  However, in order to account 
for the large EUV excess the highly evolved 
electron spectrum at the present epoch must still include sufficient
particles ahead of the cut-off.  This means that for the region
of concern, at
injection (when the power-law differential number index is
assumed to be 2.5, in accordance with our Galactic cosmic ray
index) the relativistic electron pressure
would have
exceeded that of the hot ICM by a factor of $\sim$ 4, leading to
a major confinement problem for the hot gas.  The inclusion of
cosmic ray protons exaggerates the difficulty, as
protons carry 10 - 100 times more pressure than electrons.
Thus the only other viable alternative, viz. the
originally proposed
thermal (warm) gas scenario (Lieu et al. 1996), 
must now be considered seriously.
This is especially so in the light of the recent constraints
on cosmological parameters, which point to the existence of
a warm and massive baryonic component (Cen \& Ostriker 1999,
Maloney \& Bland-Hawthorn 1999).

Turning to the core of the cluster, one sees in Fig. 1 
that the EUV is absorbed.  For
more details, in Fig. 3 is shown an image of the 
brightness of the EUV excess.
The data suggest an intermixed model (Jacobsen \& Kahn 1986)
of the ICM: the lack of soft excess 
at small radii is due entirely to
the larger amount of cold absorbing matter collected in this region.  
Our inference of the state of the ICM is reinforced by the behavior of the
SERT: the trend depicted in Fig. 1 follows a simple parametric
profile which applies equally satisfactorily to the absorption
and soft excess regions, 
with no change of behavior at the transition radius of $\sim$ 6 arcmin.
In fact, there is no particular significance in this radius (it is
much larger than the cooling flow radius 
of $\sim$ 2 arcmin; Siddiqui et al. 1998).
The observation is naturally interpreted as the combined effect of 
clumped emission regions containing a warm
component, absorbed by clouds of cold gas (neutral H
 at T $\sim 10^4$ K) in between them (statistical
quality of the data can not well constrain the size of the clouds.
However, if pressure equipartition with the hot gas is assumed,
the HI mass estimates derived in the following convert to
 an approximate cloud radius of 1-10 kpc).  
Both the warm and the cold gas distributions are
cluster-wide and centrally condensed, but with increasing radius 
the lines-of-sight are more
transparent to EUV photons created at locations along them.
For comparable intrinsic emission profiles of the soft
excess and the hot ICM, the
result is an outwardly rising SERT.

The argument for an intermixed ICM also rests upon
direct evidence for the presence of
soft excess even in the absorbed regions.  We show in Fig. 4
a core spectrum, where it can be seen that by the time intrinsic
absorption accounts for the
EUV decrement, an excess is seen in soft X-rays
(which are less absorbed).  This clearly
indicates a complex ICM where the various gas phases co-exist.  The
apparent negative soft excess within the absorption radius is
simply due to an abundance of cold clouds masking EUV emissions from
the warm and hot components.

The thermal origin of the EUV is compelling for another reason:
the widespread absorption reported here implies the existence of 
a cold phase in the midst of the well known hot phase, and then the question
naturally arises concerning why a warm phase is absent, and
is not the cause of the soft excess.  At the very least, mixing
layers on the surface of the cold clouds would suffice to produce
the intermediate phase (Fabian 1997).

The mass budgets of the three ICM components in consideration
are estimated as follows.  The intrinsic HI column density as
inferred from the central EUV absorption converts to
a density of cold clouds of $\sim$ 5 $\times$ 10$^{10}$
M$_\odot$ Mpc$^{-3}$.
This gives a mass ratio of 1:2000 between
the cold and hot gas along the line-of-sight.  Any estimate of
the mass of warm gas at the centre is likely to be inaccurate,
since the soft emission is significantly absorbed.  We therefore
considered the 12 - 15 arcmin region where this complication is
not as severe as in the centre.
The extreme softness of the
emission (Fig. 2) limits the gas temperature to kT $<$ 100 eV
(or T $<$ 10$^6$ K), with a correspondingly large mass estimate
of 1.25 $\pm^{0.4}_{0.9} \times$ 10$^{14}$ M$_\odot$ \footnote{Errors
are estimated from the statistical uncertainties in the determination of the 
warm gas emission
measure, as obtained by spectral modelling with
XSPEC 10.0.},
under the assumption of 100 \% filling factor of the
warm gas in the spherical shell of concern;
this is $\sim$ 43 $\pm^{13}_{29}$ times more massive than the hot ICM
in this region, as obtained from the hot gas density of
Siddiqui et al. (1998).  
The 1-$\sigma$ lower 
limit ratio implies $\sim$ 3 times more missing
baryons than expected (Cen \& Ostriker 1999), 
although it must be emphasized that
both the mass and bolometric luminosity can be substantially
reduced if the gas turns out to be warmer.  This can be realized
by adopting alternative emission models for the warm phase,
especially those which involve
an underionized gas, since the EUV emission efficiency is then
enhanced, and the gas can be warmer than the above temperature constraint.
Plasma in such an ionization state is easily produced by mixing
layers or shock heating.

\newpage 

\noindent
{\bf References}

\noindent
Balucinska-Church M., McCammon, D. 1992, {ApJ} {400}, 699. \\
\noindent
Bowyer S., Malina R.F. 1991, in {EUV Astronomy}, \\
\indent eds. R.F. Malina, S. Bowyer (new York:Pergamon), 397.\\
\noindent
Cen R., Ostriker J.P. 1999, {ApJ} {514}, 1-6. \\
\noindent
Maloney P.R., Bland-Hawthorn J. 1999, {ApJ} {522}, L81-84. \\
\noindent
Ensslin T.A., Biermann, P.L. 1998, {A\&A} {330},
90--98. \\
\noindent
Fabian A.C. 1997, {Science} {275}, 48--49. \\
\noindent
Jakobsen P., Kahn, S.M. 1986, {ApJ} {309}, 682 .\\
\noindent
Kaastra J.S. 1992 in An X-Ray Spectral Code for Optically Thin Plasmas \\
\indent
(Internal SRON-Leiden Report, updated version 2.0). \\
\noindent
Kaastra J.S., Lieu R.,
 Mittaz, J.P.D. et al.
 1999, {ApJ} {519}, L119. \\
\noindent
 Lieu R., Mittaz J.P.D., Bowyer S. et al. 
 1996, {Science}, {274},1335--1338. \\
\noindent
Lieu R., Bonamente M., Mittaz  
 1999a, {ApJ} {517}, L91. \\
\noindent
Lieu R., Bonamente M., Mittaz et al. 
 1999b, {ApJ} {527}, L77. \\
\noindent
 Mewe R., Gronenschild E.H.B.M., van den Oord G.H.J. 1985, {A\&A} {62}, 197 .\\
\noindent
 Mewe R., Lemen J.R., van den Oord G.H.J. 1986, {A\&A} {65}, 511.\\
\noindent
 Morrison R., McCammon, D. 1983, {ApJ} {270}, 119.\\
\noindent
 Sarazin C.L., Lieu R. 1998, {ApJ} {494}, L177. \\
\noindent
Siddiqui H., Stewart G.C., Johnston, R.M. 1998,
{A\&A}, {334}, 71.\\
\noindent
Snowden S.L., McCammon D., Burrows D.N., Mendenhall, J.A. 1994, \\
\indent {ApJ} {424}, 714.\\

\newpage

\begin{figure*}[h]
\resizebox{12cm}{!}{\rotatebox{90}{\includegraphics{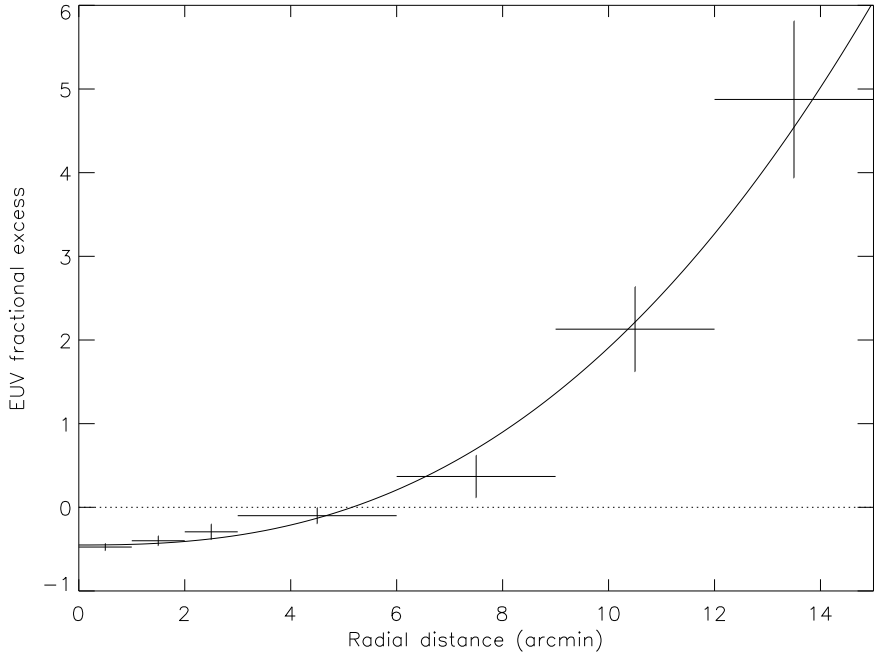}}}
\hfill
\caption{
The SERT effect illustrated by a plot against cluster
radius $r$ of the EUV fractional excess $\eta$, defined as $\eta = (p - q)/q$,
where $p$ is the DS Lex/B observed signal and $q$ is the expected
EUV emission from the hot ICM ($\eta$=0 is expected
if no absorption/soft-excess emission is present).
$q$ is determined from the best model of
the PSPC data (single temperature fits were found to be
satisfactory at all radii) with details of Galactic absorption
as quoted in the text.  The data follow a parametric profile
$\eta \propto r^{2.5}$ (solid line).
Similar results were found in LBM99 (see Fig. 1b therein, first EUVE
observation of the cluster with Galactic absorption cross section
from Balucinska-Church \& McCammon 1992) and in
L99 (EUVE reobservation, see Fig. 5b therein), 
where the DS Lex/B-to-PSPC R2 band 
count rate ratio was plotted instead. The PSPC R2 band (see Snowden et al.
1994 for definition) count rate relates
to $q$ via a radially dependent numerical factor, typically
in the vicinity of a few $\times 10^{-2}$, which scales the R2 band emissivity
of the hot gas to that in the DS Lex/B passband.
}
\end{figure*}

\begin{figure*}
\resizebox{16cm}{!}{\rotatebox{90}{\includegraphics{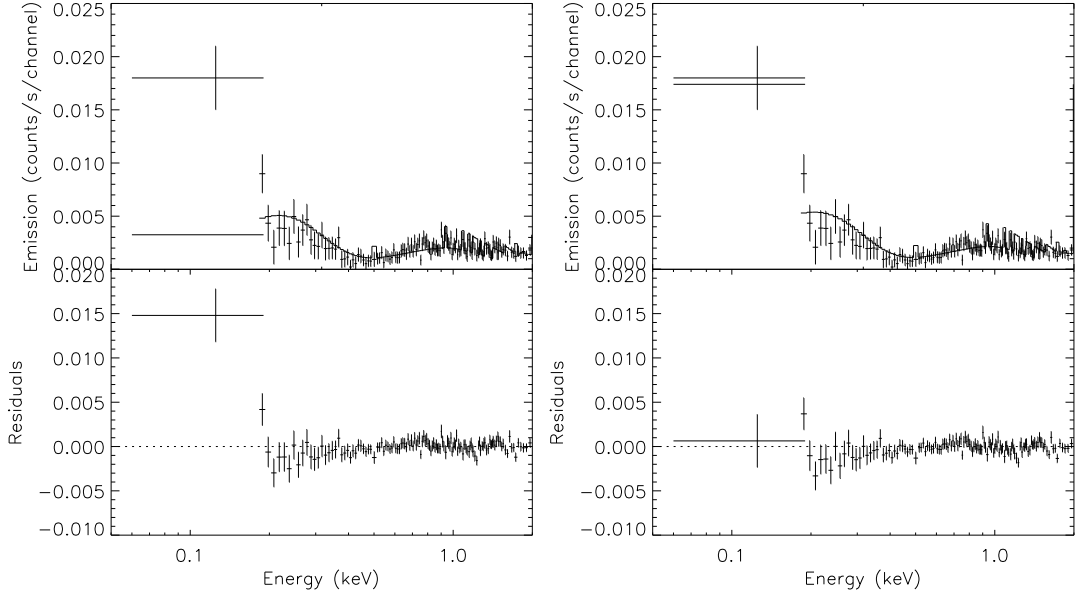}}}
\caption{
Emission models (solid line)
used to simultaneously fit the EUVE/DS
and ROSAT/PSPC data of the 12 - 15 arcmin annular region of A2199.
{\it Left Panel:} isothermal thin plasma spectrum (Mewe
et al. 1985; Mewe et al. 1986, Kaastra 1992)
at kT = 4.08 keV (Siddiqui et al. 1998) and an abundance of 0.3
 solar.  Note
the strong EUV excess recorded by the DS (left most data point)
which is not seen in soft X-rays by the PSPC (remaining data points).
{\it Right Panel:} same as the previous model, except with an
additional non-thermal component due to the IC/CMB effect (see text).
The electron population (assumed
to have an initial differential number index of 2.5, similar to
that of Galactic cosmic rays) is
$\sim$ 3.5 Gyr old, as during this period the IC/CMB and
synchrotron losses would have secured the necessary high energy
cut-off to avoid emissions in the PSPC
passband.  At the present
epoch the electron pressure is $\sim$ 25 \% that of the
hot ICM, while the initial value of this ratio was $\sim$ 400 \%.
}
\end{figure*}

\begin{figure*}
\resizebox{16cm}{!}{\includegraphics{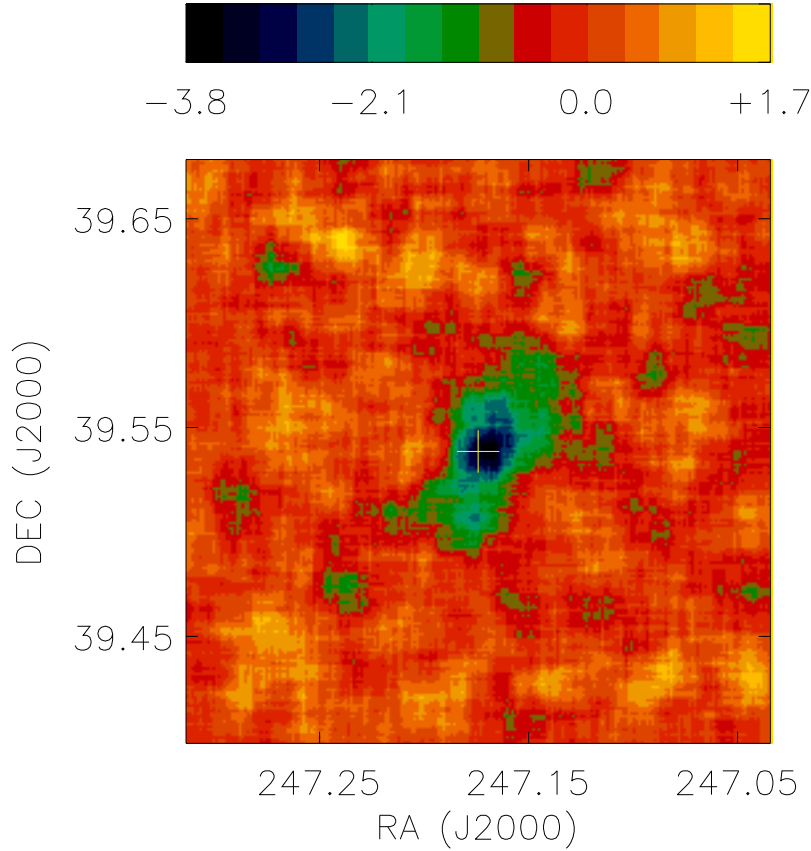}}
\vspace{-8cm}
\caption{
An image of the surface brightness of EUV excess 
($p(x,y)-q(x,y)$, using the notation of Fig.1) for
the central region of Abell 2199, obtained after subtraction
of background and contributions from the hot ICM emission ($q(x,y)$, see
text).  The pixel units (color coded) are in photons arcmin$^{-2}$ s$^{-1}$. 
Pixels of negative excess correspond to areas where the
EUV from warm and hot components are absorbed by a cold
component.  The common centroid of the cluster EUV and soft X-ray
emissions is marked by a cross.
In LBM99 (see Fig.2 bottom therein) we showed a contour plot
of the DS Lex/B-to-PSPC R2 band ratio. Given that
the hot ICM contribution is a significant fraction of the soft fluxes,
the present plot (with the hot ICM contribution removed) 
can highlight fainter EUV absorption/excess emission features.
}
\end{figure*}

\begin{figure*}
\resizebox{16cm}{!}{\rotatebox{90}{\includegraphics{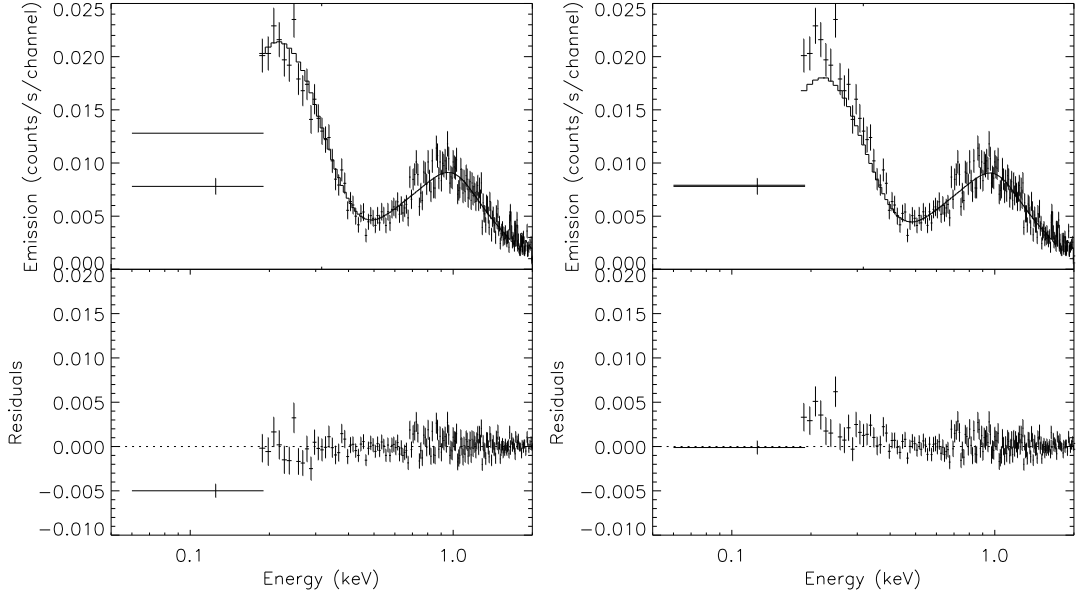}}}
\caption{
Data are as in Fig. 2, except for the 1 -- 2 arcmin
radius of A2199.  {\it Left Panel:} single temperature
emission model (kT = 3.58 keV, abundance = 0.56
solar) showing
the EUV signal in absorption.  {\it Right Panel:} Plasma properties
as above, with an intrinsic cold gas of line-of-sight HI column density 
$N_H$ = 2.7 $\times$ 10$^{19}$
cm$^{-2}$ invoked to account for the depleted EUV flux.  Note this
correction revealS
a soft X-ray excess in the PSPC 1/4- keV band, thus clearly indicating
the presence of an underlying warm component which is masked by
the cold absorbing phase.
}
\end{figure*}

\newpage

\end{document}